\documentclass[prb,reprint,superscriptaddress,
amsmath,amssymb,aps,]{revtex4-2}
\usepackage{array}
\usepackage{amsmath}
\usepackage{graphicx}
\usepackage{dcolumn}
\usepackage{bm}
\usepackage{float}
\usepackage{hyperref}
\hypersetup{colorlinks,linkcolor={blue},citecolor={blue},urlcolor={blue}}  
\usepackage{physics}
\usepackage{subcaption}
\usepackage[compatibility=false]{caption}

\usepackage{xtab,afterpage}
\usepackage[utf8]{inputenc}
\DeclareUnicodeCharacter{03C3}{\ensuremath{\sigma}}
\makeatletter
\def\LT@LR@e{\LTleft\z@   \LTright\z@}%
\makeatother

\usepackage[utf8]{inputenc}
\usepackage{textcomp}
\DeclareUnicodeCharacter{2212}{\textminus}

\begin{document}

\preprint{APS/123-QED}

\title{Real-Time Iteration Scheme for Dynamical Mean-Field Theory: A Framework for Near-Term Quantum Simulation}
\author{Chakradhar Rangi}
\affiliation{Department of Physics and Astronomy, Louisiana State University, Baton Rouge, LA 70803, USA}

\author{Aadi Singh}
\affiliation{Department of Physics and Astronomy, Louisiana State University, Baton Rouge, LA 70803, USA}
\affiliation{Center for Computation and Technology, Louisiana State University, Baton Rouge, LA 70803, USA}

\author{Ka-Ming Tam}
\affiliation{Department of Physics and Astronomy, Louisiana State University, Baton Rouge, LA 70803, USA}
\affiliation{Center for Computation and Technology, Louisiana State University, Baton Rouge, LA 70803, USA}

\date{\today}
\begin{abstract}

We present a time-domain iteration scheme for solving the Dynamical Mean-Field Theory (DMFT) self-consistent equations using retarded Green's functions in real time. Unlike conventional DMFT approaches that operate in imaginary time or frequency space, our scheme operates directly with real-time quantities. This makes it particularly suitable for near-term quantum computing hardware with limited Hilbert spaces, where real-time propagation can be efficiently implemented via Trotterization or variational quantum algorithms. We map the effective impurity problem to a finite one-dimensional chain with a small number of bath sites, solved via exact diagonalization as a proof-of-concept. The hybridization function is iteratively updated through time-domain fitting until self-consistency. We demonstrate stable convergence across a wide range of interaction strengths for the half-filled Hubbard model on a Bethe lattice, successfully capturing the metal-to-insulator transition. Despite using limited time resolution and a minimal bath discretization, the spectral functions clearly exhibit the emergence of Hubbard bands and the suppression of spectral weight at the Fermi level as interaction strength increases. This overcomes major limitations of two-site DMFT approximations by delivering detailed spectral features while preserving efficiency and compatibility with quantum computing platforms through real-time dynamics.

\end{abstract}\maketitle

\section{Introduction}

The advent of near-term quantum computing hardware promises transformative advances in simulating complex quantum many-body systems, a domain where classical methods often hit a hard ceiling \cite{Preskill2018, Daley2022, Ayral2023, Fauseweh2024, Clinton2024}. While classical algorithms struggle with the exponential growth of the Hilbert space or the fermion sign problem in Monte Carlo methods, quantum hardware offers a fundamentally different path \cite{Bauer_etal_2023,Banuls_etal_2020}. Yet, despite progress in algorithms for small molecules and optimization, strongly correlated condensed matter systems, exemplified by the deceptively simple Hubbard model, remain a formidable challenge \cite{Kanamori_1963, Hubbard_I_1963, Stanisic2022, Mondaini2022}. Accurately treating these systems requires capturing the delicate interplay between kinetic energy and strong on-site Coulomb interactions, a task that has historically defied exact solution. 

Dynamical Mean-Field Theory (DMFT) stands out as a powerful, scalable approximation to address this complexity \cite{Georges_Kotliar_1992,DMFT_RMP,Muller_Hartmann_1989a,Muller_Hartmann_1989b,Metzner_Vollhardt_1989,Jarrell_1992}. By simplifying the lattice problem into a single impurity embedded in a self-consistently determined dynamical bath, DMFT captures the essential physics of the infinite-dimensional limit. It has become a cornerstone of modern condensed matter physics, providing reliable insights into phase diagrams and the metal-to-insulator transition while incorporating dynamical quantum fluctuations that static mean-field approaches miss \cite{Anisimov_Zaanen_Andersen_1991, Kotliar2006, Held2007, Vollhardt_2011}.

Despite this success, a major challenge lies in solving the effective impurity problem to obtain the impurity Green's function accurately. Over the past three decades, a wide range of classical impurity solvers has been developed, including perturbative approaches such as Iterative Perturbation Theory (IPT) \cite{Kajueter_Kotliar_1996}, Fluctuation Exchange Approximation (FLEX) \cite{Bickers_Scalapino_White_1989}, non-crossing approximation (NCA) and its generalizations \cite{Kotliar_Ruckenstein_1986,Haule_etal_2001,Grewe_Keiter_1981, Keiter_Kimball_1971, Kirchner_Kroha_2002}; numerically exact Hamiltonian-based methods like exact diagonalization (ED) \cite{Caffarel_Krauth_1994,Capone_etal_2007,Liebsch_Ishida_2011,Kuramoto_1983}, numerical renormalization group (NRG) \cite{Bulla_Costi_Pruschke_2008,Peters_etal_2006}, density matrix renormalization group (DMRG) \cite{Ganahl_etal_2015,Wolf_etal_2015}, and Fork Tensor Product States \cite{Bauernfeind_etal_2017,Weh_etal_2021}; as well as action-based continuous-time quantum Monte Carlo (CT-QMC) variants and tensor-train approximations \cite{CTQMC_RMP,Rubtsov_Savkin_Lichtenstein_2005,Haule_2007,Assaad_Lang_2007,Gull_etal_2008,Werner_etal_2006, Erpenbeck_etal_2023}. More recent directions include machine-learning-inspired models that replace traditional solvers for the single-band Hubbard model \cite{Arsenault_etal_2014,Rigo_etal_2020,Sheridan_etal_2021,Walker_etal_2022,Sturm_etal_2021}. While these methods provide high accuracy for the single-orbital impurity problem, most operate in imaginary time or frequency space and scale poorly with increasing bath size or interaction strength on constrained hardware.

With advances in quantum computing, there is growing interest in solving the impurity problem and the full lattice Hubbard model using quantum algorithms \cite{PhysRevA.92.062318,PhysRevX.6.031045}. Unlike most quantum chemistry applications, DMFT requires the full Green's function, while near-term (Noisy Intermediate-Scale Quantum) devices offer limited qubits and gate counts. Pioneering efforts have showed that it is possible to find the Green's function for the single impurity Anderson model using the Variational Quantum Eigensolver (VQE) as the ground state  \cite{Ayral2025,Baul_etal_2025,kreula2016few,Steckmann_etal_2023,Besserve_Ayral_2022,Kreula_Clark_Jaksch_2016,Rungger_etal_2019,Jaderberg_etal_2020,Yao_etal_2021,Steckmann_etal_2023,Jamet_etal_2021,Jamet_Agarwal_Rungger_2022,Ehrlich_Urban_Elsasser_2024,Jamet_etal_2025,Baul_etal_2025,Selisko_etal_2024,Jones_etal_2025,Dhawan_Zigid_Motta_2023,Karabin2026,Perez_etal_2025}. These groundbreaking proof-of-concept demonstrations on superconducting and trapped-ion platforms capture qualitative features like the Mott transition, but suffer from severe limitations: the minimal bath reduces rich dynamics to a single (or few) fitting parameter(s), failing to reproduce detailed spectral features such as distinct Hubbard bands or accurate quasi-particle weights essential for real materials.

To overcome these limitations and build a scalable pipeline for quantum simulation, we study a time-domain iteration scheme for the DMFT self-consistent equations. Unlike conventional classical DMFT solvers that predominantly rely on frequency-space formalisms, we formulate our approach directly in terms of real-time retarded Green’s functions. We map the effective impurity problem onto a finite one-dimensional chain with a small number of bath sites (5 sites), solved via exact diagonalization as a proof-of-concept. The hybridization function is updated iteratively by fitting the retarded Green's function in real time, bypassing the numerically unstable inversion between time and frequency domains. This approach is specifically tailored for quantum implementation: it enables the simulation of larger effective baths while maintaining minimal qubit and gate depth requirements, respecting the constraints of near-term devices. We demonstrate stable convergence across a wide range of interaction strengths for the half-filled Hubbard model on a Bethe lattice. Despite using limited time resolution and a minimal bath discretization, our results clearly exhibit the emergence of Hubbard bands and the suppression of spectral weight at the Fermi level, features that are inaccessible to the two-site approximation. This work establishes a robust framework for capturing key spectral features via DMFT simulations on near-term quantum devices.

The structure of this paper is as follows: In Section II,  we present the Hubbard model and derive the DMFT self-consistency loop for the Bethe lattice. Readers familiar with DMFT can skip this section. Section III details the algorithm we propose to solve the DMFT self-consistent equations using a small quantum cluster size with a real-time single-particle Green's function. In Section IV, we present the results from the weakly interacting correlated metal limit to the strongly interacting Mott insulator limit. Section V concludes with a summary of our findings and potential directions for future research.

\section{Model and method}

\subsection{Hubbard Model and Dynamical Mean Field Theory}\label{Subsection:Model}

We consider the Hubbard model on a lattice defined by the Hamiltonian
\begin{gather}
H = -\sum_{\langle i,j\rangle,\sigma} t_{ij} \left( c_{i,\sigma}^\dagger c_{j,\sigma} +c_{j,\sigma}^\dagger c_{i,\sigma} \right) - \mu\sum_{j,\sigma} n_{j\sigma} \nonumber\\
+ U \sum_j n_{j\uparrow} n_{j\downarrow},
\label{eq:hamiltonian}
\end{gather}
where the summation $\langle i,j\rangle$ runs over nearest-neighbor pairs, $c_{j,\sigma}^\dagger$ and $c_{j,\sigma}$ are the fermionic creation and annihilation operators at site $j$ with spin $\sigma = \uparrow, \downarrow$, and $n_{j\sigma} = c_{j,\sigma}^\dagger c_{j,\sigma}$ is the number operator. The parameter $t_{ij}$ represents the hopping amplitude, $\mu$ is the chemical potential, and  $U$ represents the on-site Hubbard interaction strength, capturing strong electron correlations. We assume symmetric hopping ($t_{ij}=t_{ji}$) and set $\mu=0$ to ensure half-filling. In the following, we define the scaled hopping $t^*$ as our unit of energy ($t^*=1$) to distinguish it from the time variable t.

\subsection{DMFT}\label{Subsection:Method}
 The core idea of DMFT is to consider the model in the infinite-dimensional limit so that the self-energy becomes space-independent. The self-energy is then computed by mapping the lattice problem onto a single-site impurity problem coupled to a self-consistently determined dynamical bath \cite{DMFT_RMP,Muller_Hartmann_1989a,Muller_Hartmann_1989b,Metzner_Vollhardt_1989}. We first consider the original model on a lattice using the path integral formalism.
The partition function at inverse temperature ($\beta$) using Grassmann fields $(\bar{\psi},\psi)$ is given as 
\begin{equation}
\mathcal{Z} = \int \prod_{i} \mathcal{D}[\bar{\psi}_{i\sigma}, \psi_{i\sigma}] e^{-\mathcal{S}[\bar{\psi},\psi]},
\end{equation}
where the action $\mathcal{S}$ is given by:
\begin{align}
\mathcal{S} = &\sum_{\sigma} \int_{0}^{\beta} \int_{0}^{\beta} d\tau d\tau'\sum_{i,j} \bar{\psi}_{i\sigma}(\tau) [G_{0}^{-1}]_{ij}(\tau,\tau') \psi_{j\sigma}(\tau') \nonumber \\ &+ \mathcal{S}_{U}.
\end{align}
For notational brevity, we define:
\begin{equation}
   \mathcal{S}_{U} \equiv \int_{0}^{\beta} d\tau \sum_{i} U \bar{\psi}_{i\uparrow}(\tau) \bar{\psi}_{i\downarrow}(\tau) \psi_{i\downarrow}(\tau) \psi_{i\uparrow}(\tau).
\end{equation}
The inverse of the bare lattice Green's function is given by:
\begin{equation}
[G_{0}^{-1}]_{ij}(\tau, \tau') =  \left[ (\partial_{\tau} - \mu) \delta_{ij} + t_{ij} \right]\delta(\tau - \tau').
\end{equation}
The Grassmann fields satisfy the boundary condition $\psi_{i\sigma}(0) = -\psi_{i\sigma} (\beta)$ for fermions.

We consider the Hubbard model defined on the Bethe lattice with a large coordination number ($d$). To ensure the Hamiltonian remains well-defined in the limit $d \rightarrow \infty$,  we must scale the hopping amplitude as $t_{ij} \propto1/\sqrt{d}$. We then integrate out all degrees of freedom except for a specific site, denoted as site 0; the partition function becomes:
\begin{equation}
    e^{-\mathcal{S}_{eff}(\bar{\psi}_0, \psi_0)} = \int \prod_{i \neq 0} \mathcal{D}[\bar{\psi}_i, \psi_i] e^{-\mathcal{S}}, 
\end{equation}
The resulting contribution to the effective action from the environment can be expanded in a series in powers of the hopping amplitude $t_{0j}$. In the limit $d \to \infty$, only the second-order term survives:
\begin{equation}
\Delta \mathcal{S} = -\sum_{\sigma} \int d\tau d\tau' \bar{\psi}_{0\sigma}(\tau) \Delta(\tau, \tau') \psi_{0\sigma}(\tau').
\end{equation}
Higher-order terms vanish in the limit $d\rightarrow\infty$ as they scale with $\mathcal{O}(1/d^2)$. This reduces the lattice problem to a single impurity coupled to a non-interacting bath, described by the effective action: \begin{eqnarray}&&\mathcal{Z}_{imp} = \int \mathcal{D}[\bar{\psi}_0, \psi_0] \nonumber \exp \\&&\left( -\sum_{\sigma} \int d\tau d\tau' \bar{\psi}_{0\sigma}(\tau) \mathcal{G}_{0}^{-1}(\tau, \tau') \psi_{0\sigma}(\tau') + \mathcal{S}_{0U} \right),\end{eqnarray} where the bare Green's function of the impurity is defined via its inverse, \begin{equation}
\mathcal{G}_{0}^{-1}(\tau, \tau') = \Bigl(\frac{\partial}{\partial \tau}  - \mu\Bigr) \delta(\tau - \tau') - \Delta(\tau,\tau'),
\end{equation}and the local interaction at site $0$ is given as
\begin{equation}
\mathcal{S}_{0U} = U\int_{0}^{\beta} d\tau  \bar{\psi}_{0\uparrow}(\tau) \bar{\psi}_{0\downarrow}(\tau) \psi_{0\downarrow}(\tau) \psi_{0\uparrow}(\tau).
\end{equation}

The Bethe lattice possesses a unique tree-like topology containing no closed loops. Consequently, the removal of the central site $0$ completely decouples its neighbors from one another, ensuring that the cavity Green's function (the Green's function of the lattice with site $0$ removed) is diagonal in the site indices: $G^{(0)}_{jk} = 0$ for $j \neq k$. The hybridization function is therefore defined as a sum over the $d$ independent neighbors of site $0$:

\begin{equation}
    \Delta(\tau, \tau') = \sum_{j=1}^{d} t_{0j}^2 G^{(0)}_{jj}(\tau, \tau').
\end{equation}

In the limit of infinite coordination number ($d \to \infty$), the bare hopping amplitude $t_{0j}$ must be scaled to maintain a finite kinetic energy per site. We define the scaled hopping parameter $t^*$ such that the nearest-neighbor hopping scales as $t_{0j} = t^*/\sqrt{d}$. Crucially, in this limit, the removal of a single site becomes negligible, allowing us to approximate the cavity Green's function with the full local Green's function of the lattice, i.e., $G^{(0)}_{jj}(\tau, \tau') \approx G_{loc}(\tau, \tau')$. The self-consistency condition thus simplifies to:
\begin{equation}\label{eqn:Self_consistency_delta}
    \Delta(\tau, \tau') = (t^*)^2 G_{loc}(\tau, \tau').
\end{equation}

Collecting the above results, we define the standard DMFT algorithm for the Bethe lattice at infinite coordination number:

\begin{enumerate}
    \item Set an initial guess for the hybridization function $\Delta(\tau,\tau')$.
    \item Construct the bare impurity Green's function $\mathcal{G}_0$ using the Dyson equation:
    $$ \mathcal{G}_0^{-1}(\tau, \tau') = (\partial_\tau - \mu)\delta(\tau-\tau') - \Delta(\tau, \tau'). $$
    \item Solve the effective impurity problem defined by $\mathcal{G}_0$ and the local interaction $U$ to obtain the impurity Green's function $G_{imp}(\tau, \tau')$.
    \item Update the hybridization function using the self-consistency condition (Eq.~\ref{eqn:Self_consistency_delta}):
    \begin{equation}
        \Delta_{new}(\tau, \tau') = (t^*)^2 G_{imp}(\tau, \tau').
    \end{equation}
    \item Repeat from Step 2 until convergence is attained.
\end{enumerate}

The primary computational bottleneck in this algorithm is the solution of the effective impurity problem. Unlike standard solvers that exploit imaginary-time statistical mechanics, we propose a formulation based on real-time retarded Green's functions to ensure compatibility with near-term quantum hardware. The derivation above utilizes the imaginary-time path integral formalism, where temporal integrals run from 0 to $\beta$. Moving forward, we focus exclusively on the zero temperature limit ($T\rightarrow 0$, $\beta \rightarrow \infty$). In this limit, the system settles into its ground state, and we replace the periodicity in imaginary time with unitary evolution in real time, which is native to quantum hardware. Consequently, assuming time-translational invariance in equilibrium, we replace the two-time hybridization $\Delta(\tau,\tau')$ with the single-time retarded hybridization function denoted by $\Delta^R(t)\equiv \Delta^R(t,0)$. The self-consistency condition \eqref{eqn:Self_consistency_delta} transforms directly into the real-time domain as:
\begin{equation}
 \Delta^R(t) =  (t^*)^2 G^R_{loc}(t). 
\end{equation}

\section{Time-Domain Iteration Scheme}

To satisfy the self-consistency condition derived above (Eq.~\ref{eqn:Self_consistency_delta}) on near-term quantum hardware, we must bridge the gap between the theoretical infinite bath and the practical limits of available qubits. We achieve this by mapping the effective impurity problem onto a discrete, one-dimensional quantum chain. In this work, we mimic the Anderson impurity model using a 6-site cluster composed of a single impurity ($N_{imp}=1$) coupled to a finite bath of $N_{bath}=5$ sites. We specifically select an odd number of bath sites to ensure the inclusion of a zero-energy mode, which is essential for preserving particle-hole symmetry in the bath discretization~\cite{Liebsch_Ishida_2011}.
\begin{figure}[ht] 
    \centering
    \includegraphics[width=0.5\textwidth]{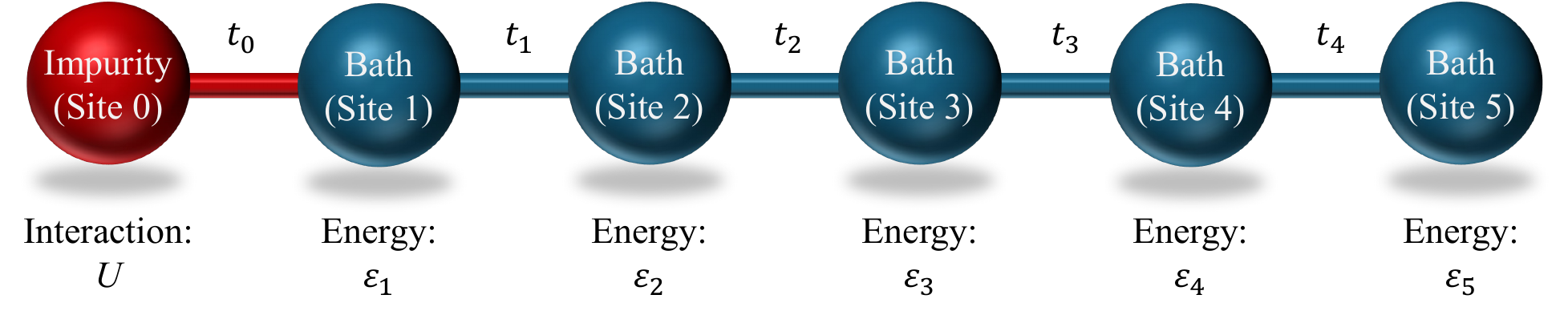}
        \caption{Schematic of the effective Anderson impurity model mapped onto a linear chain geometry. The interacting impurity (Site 0) with Coulomb repulsion $U$ is coupled to a finite, non-interacting bath of $N_{bath} = 5$	sites. The chain is parameterized by the hopping amplitudes $t_i$ and on-site energies $\varepsilon_i$.}
        \label{fig:lattice}
\end{figure}
The geometry of this effective model is illustrated in Fig.~\ref{fig:lattice}. The impurity (site 0) connects directly to the first bath site (site 1) via hopping $t_0$, while the remaining bath sites are chained via nearest-neighbor hoppings $t_i$. This architecture represents a systematic improvement over the minimal two-site DMFT approximation ($N_{bath}=1$). By increasing the number of bath sites, we introduce additional degrees of freedom that allow the discrete bath to approximate the continuous hybridization function $\Delta(t)$ with progressively higher spectral resolution. The Hamiltonian for this effective one-dimensional chain is given by:
\begin{equation}
\begin{split}
    H_{\text{chain}} = H_{imp} + H_{hyb} + H_{bath},
    \label{eqn:Chain_Hamiltonian}
\end{split}
\end{equation}
where $H_{imp} = U n_{0\uparrow}n_{0\downarrow}$ describes the interacting impurity, and $H_{hyb} = - t_0\sum_{\sigma} \left(c_{0\sigma}^\dagger c_{1\sigma} + \text{h.c.} \right)$ is the impurity-bath coupling. The bath Hamiltonian $H_{bath}$ describes the non-interacting bath sites ($i\geq 1$) and is defined as:
\begin{equation}
    H_{bath} = \sum_{\sigma} \left[ - \sum_{i=1}^{N_{b}-1} t_{i}(c_{i\sigma}^{\dagger}c_{i+1\sigma} + h.c.) + \sum_{i=1}^{N_{b}}\epsilon_{i}n_{i\sigma} \right].
\end{equation}
Here $N_b=5$ is the number of bath sites. Note that the on-site interaction $U$ is local to the impurity (site 0). The chain is fully characterized by the set of hopping amplitudes $\{t_0, t_1, t_2, t_3, t_4\}$ and on-site energies $\{\epsilon_1, \epsilon_2, \epsilon_3, \epsilon_4, \epsilon_5\}$. For the half-filled case, particle-hole symmetry imposes constraints that reduce the parameter space. This requires the hopping amplitudes to be symmetric ($t_1=t_4$, $t_2=t_3$) and the on-site potentials to be antisymmetric about the chemical potential ($\epsilon_1=-\epsilon_5$, $\epsilon_2=-\epsilon_4$, with $\epsilon_3=0$).

The goal of the iteration scheme is to determine the self-consistent hybridization function $\Delta(t)$ by iteratively fitting these chain parameters. The complete algorithm is defined as follows:

\begin{enumerate}
    \item \textbf{Initialization:}
    We first establish an initial condition for the hybridization, $\Delta^{(0)}(t)$, by solving the self-consistent equation in the non-interacting limit ($U=0$). For a Bethe lattice with infinite coordination number, the non-interacting density of states is semi-circular \cite{DMFT_RMP}:
    \begin{equation}
        \rho(\omega) = \frac{1}{2\pi {t^*}^2} \sqrt{4(t^*){^2} - \omega^2}.
    \end{equation}
    The corresponding retarded Green's function in the time domain is obtained via Fourier transform, yielding an analytical expression involving the Bessel function of the first kind ($J_1$):
    \begin{equation}
        G_{loc}^{R}(t) = -i \theta(t) \frac{J_1(2tt^*)}{tt^*}.
    \end{equation}
    We set the initial guess as $\Delta^{(0)}(t) = (t^*)^2 G_{loc}^R(t)$. Alternatively, one may use other initial guesses.

    \item \textbf{Bath Parameter Fitting:}
    At each iteration $k$, we characterize the quantum chain by finding the parameters $\mathcal{P} = \{t_i, \epsilon_i\}$ that best mimic the current retarded hybridization $\Delta^{(k)}(t)$. The hybridization of the discrete chain is given by:
    \begin{equation}
        \Delta_{chain}^R(t; \mathcal{P}) = t_0^2 g_{11}^R(t; \mathcal{P}),
    \end{equation}
    where $g_{11}^R(t)$ is the Green's function of the first bath site, calculated via exact diagonalization of the bath Hamiltonian $H_{bath}$:
    \begin{equation}
        g_{11}^{R}(t) = -i\theta(t) \sum_{m=1}^{N_b} |U_{1m}|^2 \exp(-i\lambda_m t),
    \end{equation}
    where $\lambda_m$ are the eigenvalues of $H_{bath}$ and $U_{1m}$ is the first component of the corresponding $m$-th eigenvector.
    
    We determine the optimal parameters by minimizing the error functional $\chi^2$ over the discrete time grid $\{\bar{t}_n\}$:
    \begin{equation}
        \chi^2 (\mathcal{P}) = \sum_{n} |\Delta^{(k)}(\bar{t}_n) - \Delta_{chain}^R(\bar{t}_n; \mathcal{P})|^2.
    \end{equation}

    \item \textbf{Impurity Solver:}
    With the fitted parameters, the effective impurity problem is fully specified. In the proposed hybrid framework, this step is executed on quantum hardware via two distinct simulation stages: first, preparing the ground state of the full chain Hamiltonian [Eq.\eqref{eqn:Chain_Hamiltonian}], and second, simulating the real-time evolution of the excitations. However, in this proof-of-concept study, we employ Exact Diagonalization as a proxy for a noise-free quantum computer to directly compute the interacting impurity Green's function $G_{imp}^R(t)$. To ensure $G_{imp}^R(t)$ remains well-defined and to suppress long-time errors, which are particularly relevant for Trotterized evolution on quantum hardware, we apply a damping factor $\eta$. In the time domain, this is implemented as:
    \begin{equation}
        G_{imp}^R(t) \leftarrow G_{imp}^R(t) e^{-\eta t}.
    \end{equation}
     This is equivalent to the frequency-domain shift $E \to E - i\eta$.

    \item \textbf{Self-Consistency Update:}
    We compute the new target hybridization function using the DMFT self-consistency condition:
    \begin{equation}
        \Delta^R_{new}(t) = (t^*)^2 G_{imp}^R(t).
    \end{equation}

    \item \textbf{Convergence Check:}
    Finally, we update the hybridization for the next iteration using linear mixing to ensure stability:
    \begin{equation}
        \Delta^{(k+1)}(t) = \alpha \Delta^{(k)}(t) + (1-\alpha) \Delta_{new}(t),
    \end{equation}
    where $\alpha$ is the mixing parameter. We typically set $\alpha =0.8$. The loop (Steps 2--5) repeats until convergence is attained.
\end{enumerate}

\begin{figure*}[t] 
    \begin{subfigure}[b]{0.24\textwidth} 
        \centering
        \includegraphics[width=\textwidth]{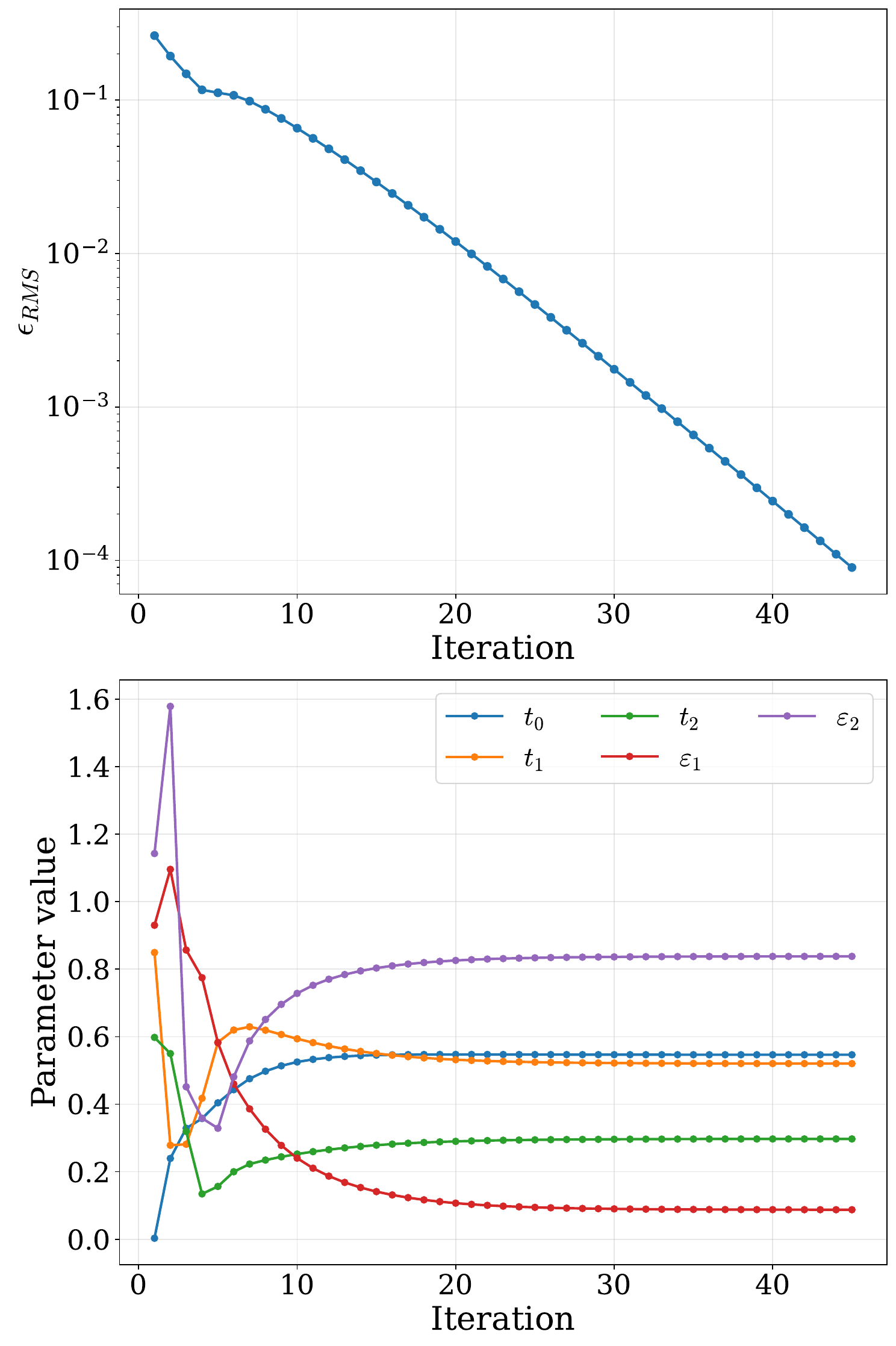}
        \caption{$U=2$}
        \label{fig:2a}
    \end{subfigure}
    \hfill 
    \begin{subfigure}[b]{0.24\textwidth}
        \centering
        \includegraphics[width=\linewidth]{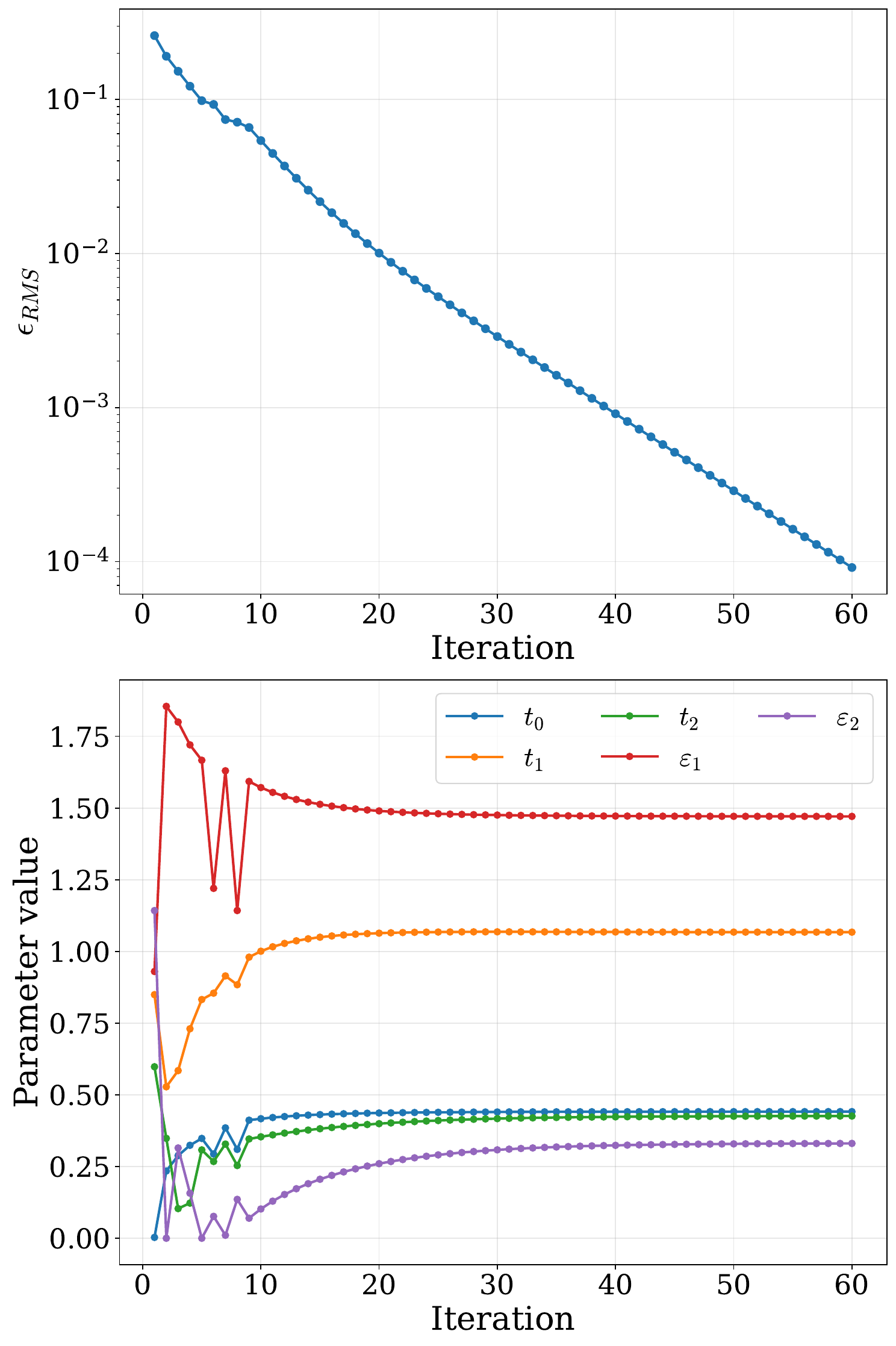}
        \caption{$U=4$}
        \label{fig:2b}
    \end{subfigure}
    \begin{subfigure}[b]{0.24\textwidth}
        \centering
        \includegraphics[width=\linewidth]{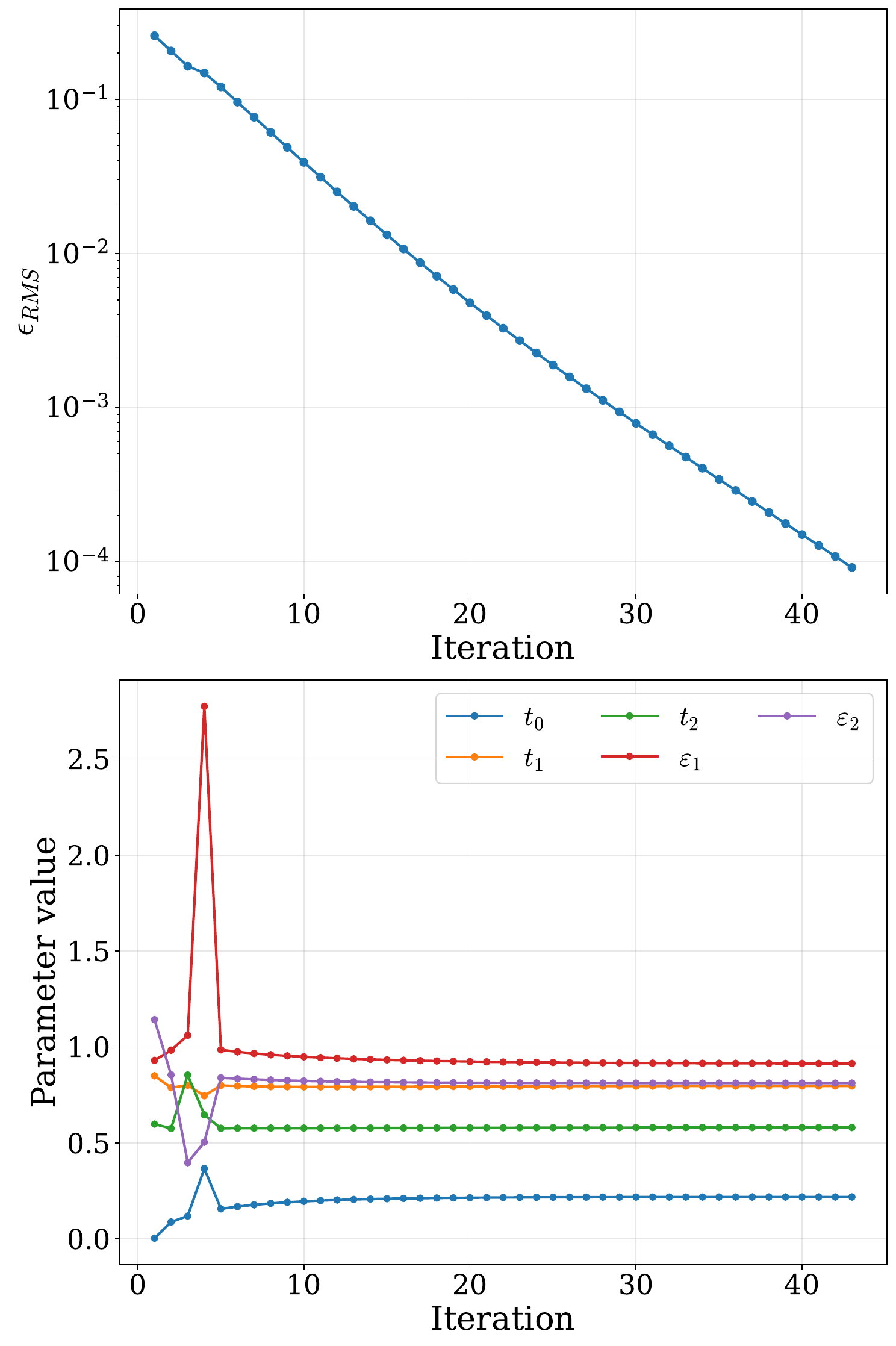}
        \caption{$U=6$}
        \label{fig:2c}
    \end{subfigure}
    \hfill
    \begin{subfigure}[b]{0.24\textwidth}
        \centering
        \includegraphics[width=\linewidth]{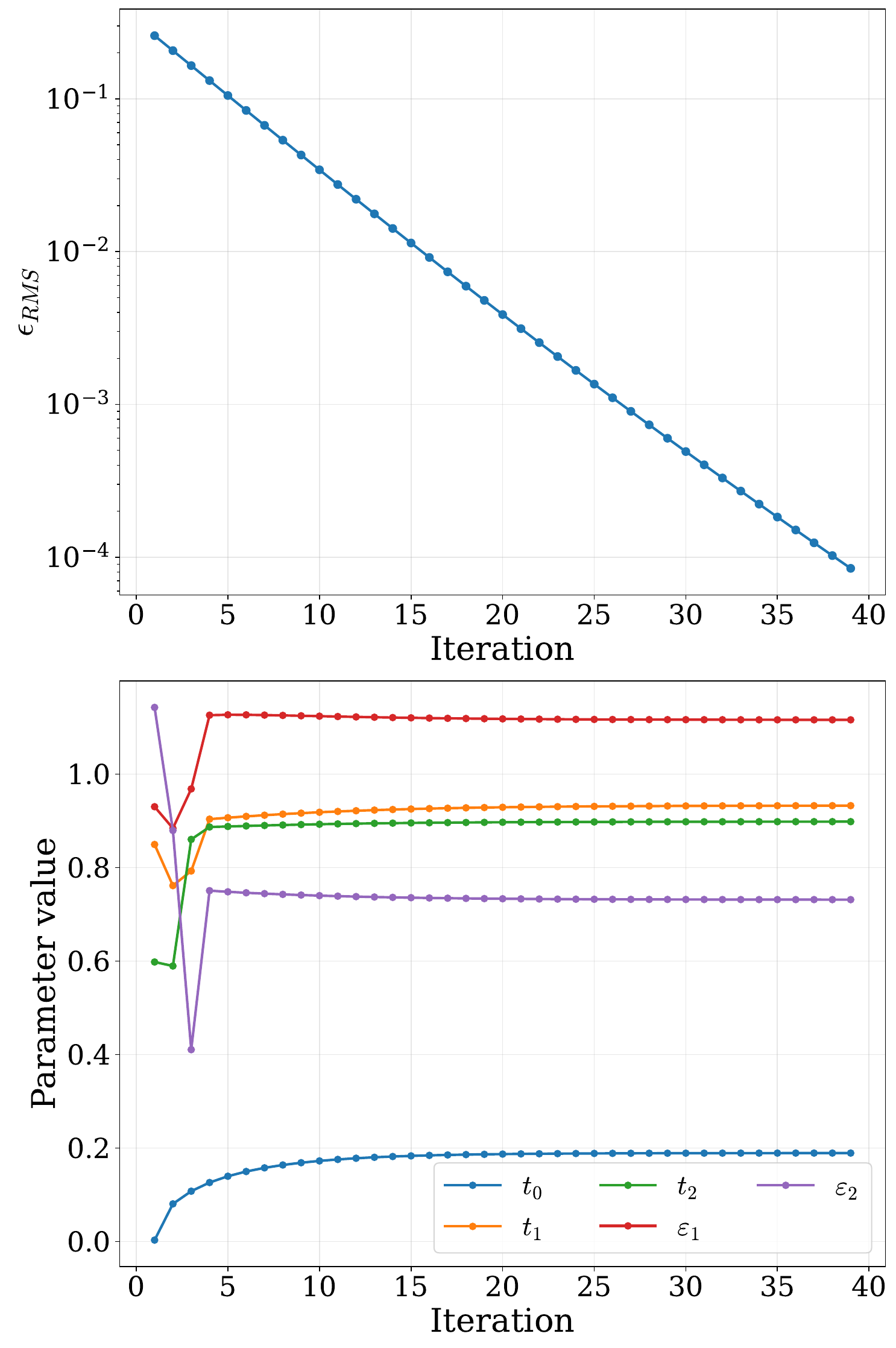}
        \caption{$U=8$}
        \label{fig:2d}
    \end{subfigure}
    
    \caption{Convergence analysis of the iterative fitting procedure across the metal-insulator crossover. The columns correspond to interaction strengths $U = 2, 4, 6,$ and $8$, respectively. \textbf{Top Row:} The convergence of the self-consistency loop, quantified by the root mean square error $\epsilon_{RMS}$	
  between consecutive hybridization functions [as defined in Eq. \eqref{eqn:RMS}], plotted against the iteration number on a logarithmic scale. \textbf{Bottom Row:} The evolution of the five independent bath parameters $\{t_0, t_1, t_2, \epsilon_1, \epsilon_2\}$ as a function of iteration count. The parameters rapidly settle to their fixed-point values within approximately 20-30 iterations, confirming the stability of the method even in the strongly correlated Mott insulating phase ($U=8$).}
    \label{fig:iteration}
\end{figure*}


\section{Results}\label{Section:Results}

In this section, we benchmark the proposed time-domain iteration scheme by solving the half-filled Hubbard model across a wide range of interaction strengths, spanning  from a weakly correlated metal to a Mott insulator. We focus specifically on paramagnetic solutions and, for notational brevity, suppress the spin index $\sigma$ in the following discussion. The numerical simulations are performed on a time domain of $t \in [0, 20)$ using a uniform grid of $N_t = 100$ points, with a damping factor set to $\eta = 0.2$. We deliberately employ this relatively coarse temporal resolution to assess the robustness of the algorithm to determine whether such restricted time-domain resolution is sufficient to capture the critical spectral features of the metal-to-insulator transition, particularly the formation of the Hubbard bands.

\subsection{Convergence Analysis}
We first evaluate the stability of the iteration scheme across distinct interaction regimes. Figure~\ref{fig:iteration} illustrates the convergence behavior for four representative interaction strengths: $U = 2, 4, 6,$ and $8$, corresponding to the weakly correlated metal, moderately correlated metal, critical region near the metal-insulator transition, and the Mott insulating regime, respectively. The upper panels of Fig.~\ref{fig:iteration} track the root mean square error of the hybridization function between consecutive iterations $k$ and $k-1$, defined as:
\begin{equation}
    \epsilon_{RMS} = \sqrt{ \frac{1}{N_t} \sum_{i=1}^{N_t} |\Delta^{(k)}(t_i) - \Delta^{(k-1)}(t_i)|^2 }. \label{eqn:RMS}
\end{equation} We observe that convergence is consistently achieved within approximately 30 iterations across all interaction strengths, demonstrating the robustness of our approach. The convergence rate appears to be relatively uniform across different values of $U$, which is particularly encouraging as it suggests that the method does not encounter significant stability issues even in the strongly correlated Mott insulating regime where $U = 8$.

The lower panels of Fig.~\ref{fig:iteration} display the evolution of the five independent bath parameters $\{t_0, t_1, t_2, \epsilon_1, \epsilon_2\}$ as functions of the iteration count. Recall that particle-hole symmetry imposes the constraints $t_1=t_4$, $t_2=t_3$, $\epsilon_1=-\epsilon_5$, $\epsilon_2=-\epsilon_4$, and $\epsilon_3=0$.
For weak interactions ($U = 2$) in Fig. \ref{fig:2a}, we observe that the parameters converge smoothly with minimal oscillations. The hopping parameter $t_0$ settles to a relatively large value, indicating strong coupling between the impurity and the bath, which is characteristic of a metallic state. The on-site energies remain small, suggesting that the bath states remain broadly distributed around the Fermi level to mimic the metallic continuum. 

\begin{figure*}[t] 
    \centering
    \begin{subfigure}[b]{0.24\textwidth} 
        \centering
        \includegraphics[width=\textwidth]{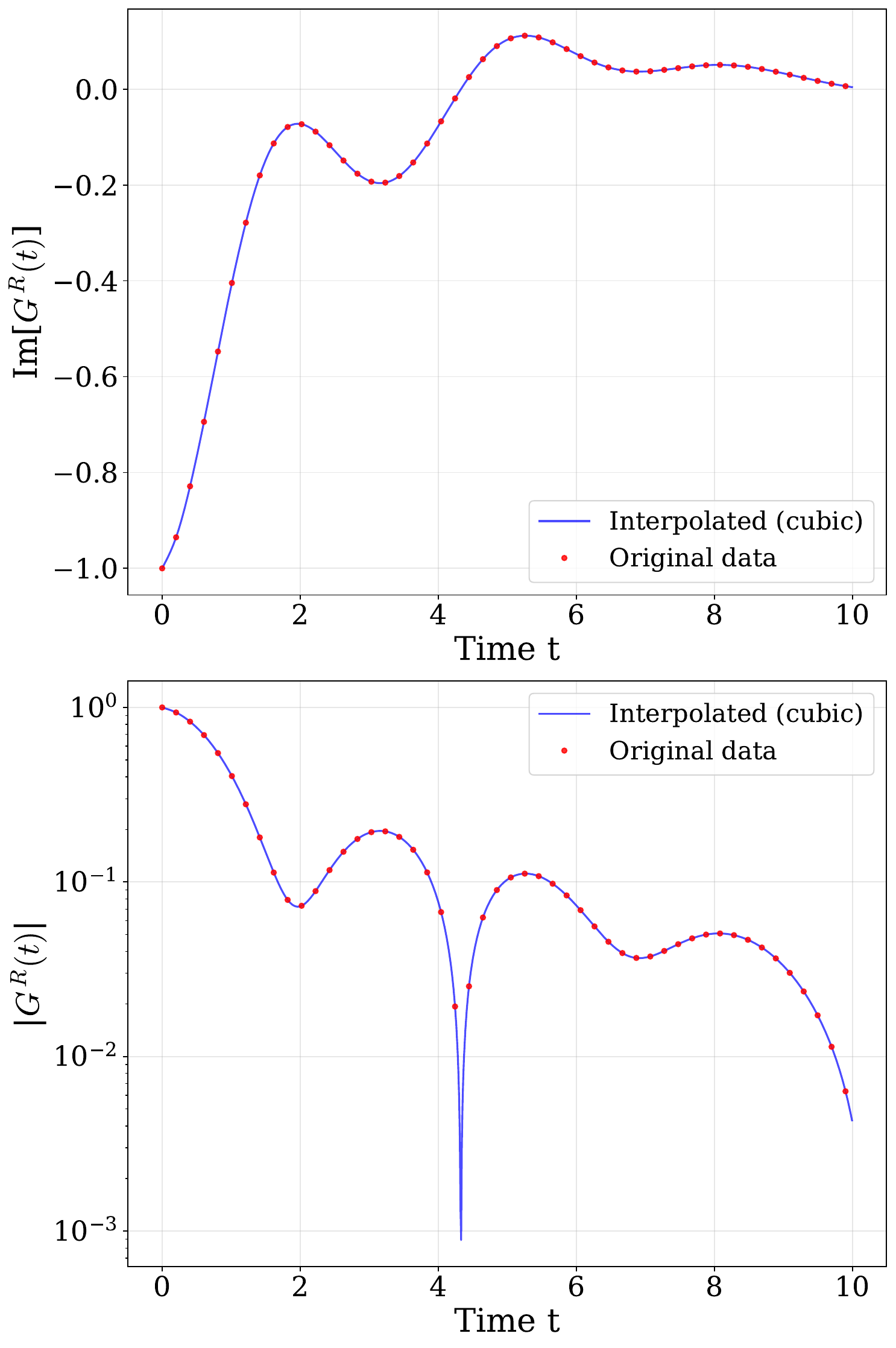}
        \caption{$U=2$}
        \label{fig:3a}
    \end{subfigure}
    \hfill 
    \begin{subfigure}[b]{0.24\textwidth}
        \centering
        \includegraphics[width=\linewidth]{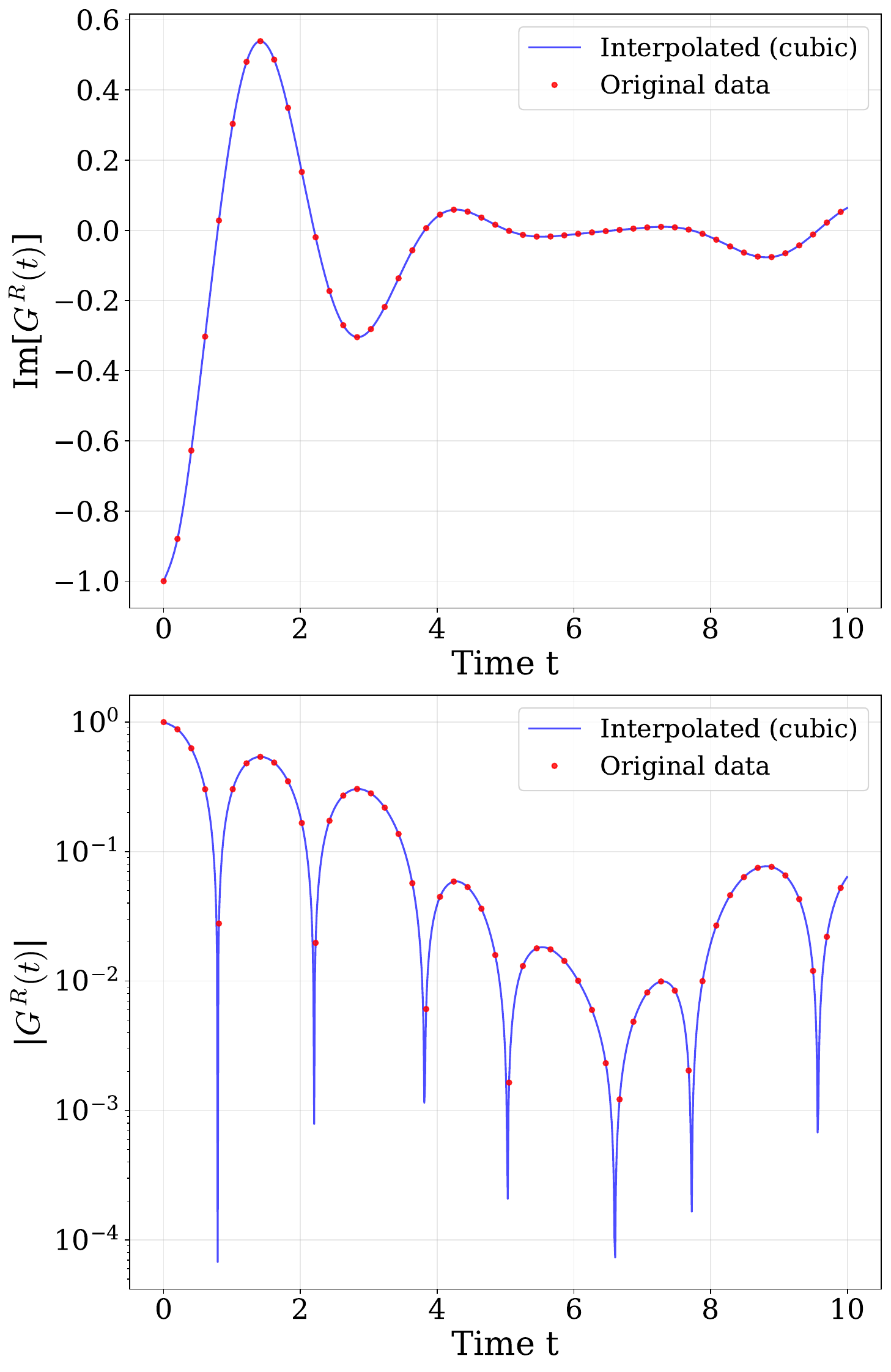}
        \caption{$U=4$}
        \label{fig:3b}
    \end{subfigure}
    \begin{subfigure}[b]{0.24\textwidth}
        \centering
        \includegraphics[width=\linewidth]{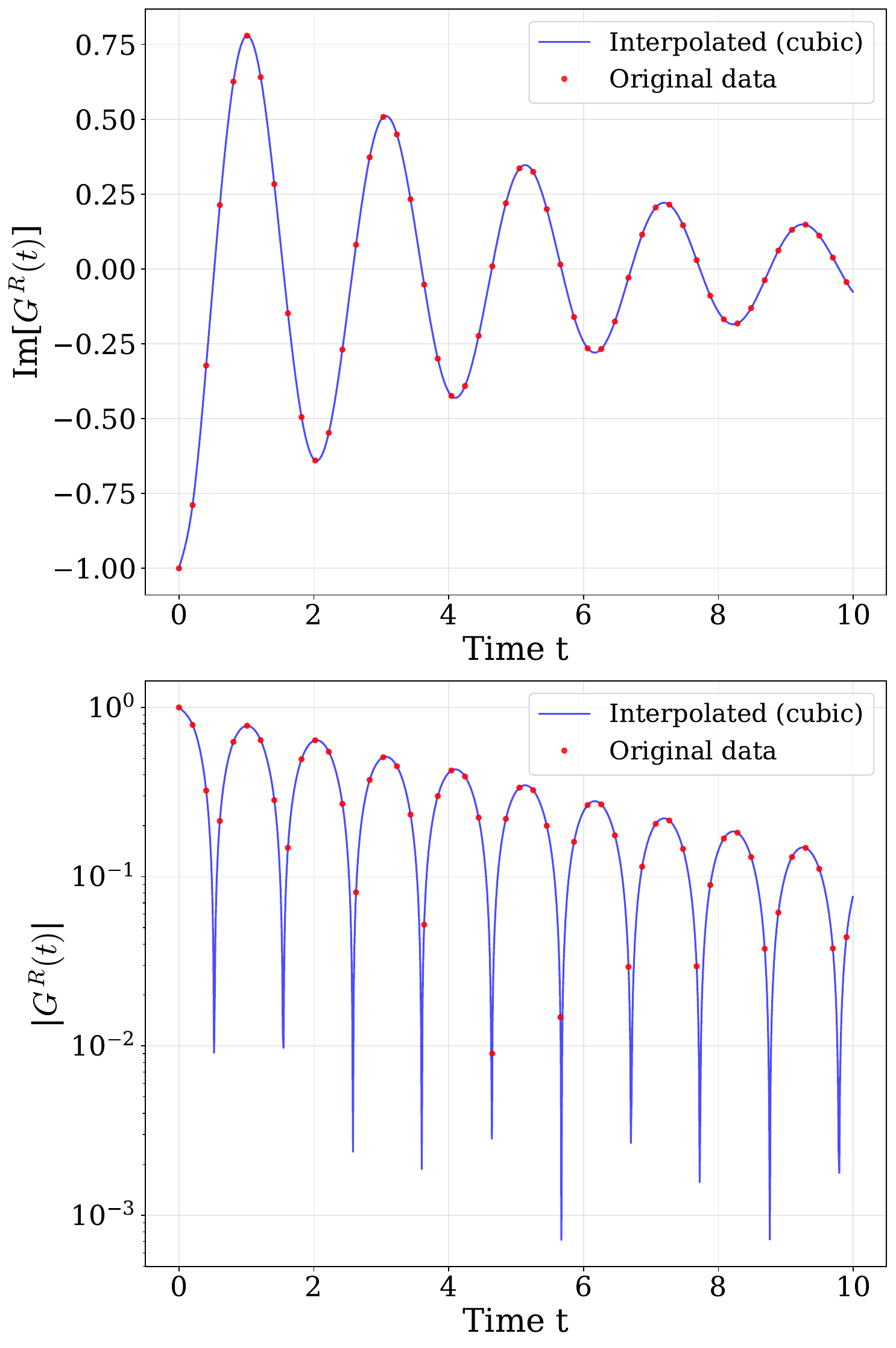}
        \caption{$U=6$}
        \label{fig:3c}
    \end{subfigure}
    \hfill
    \begin{subfigure}[b]{0.24\textwidth}
        \centering
        \includegraphics[width=\linewidth]{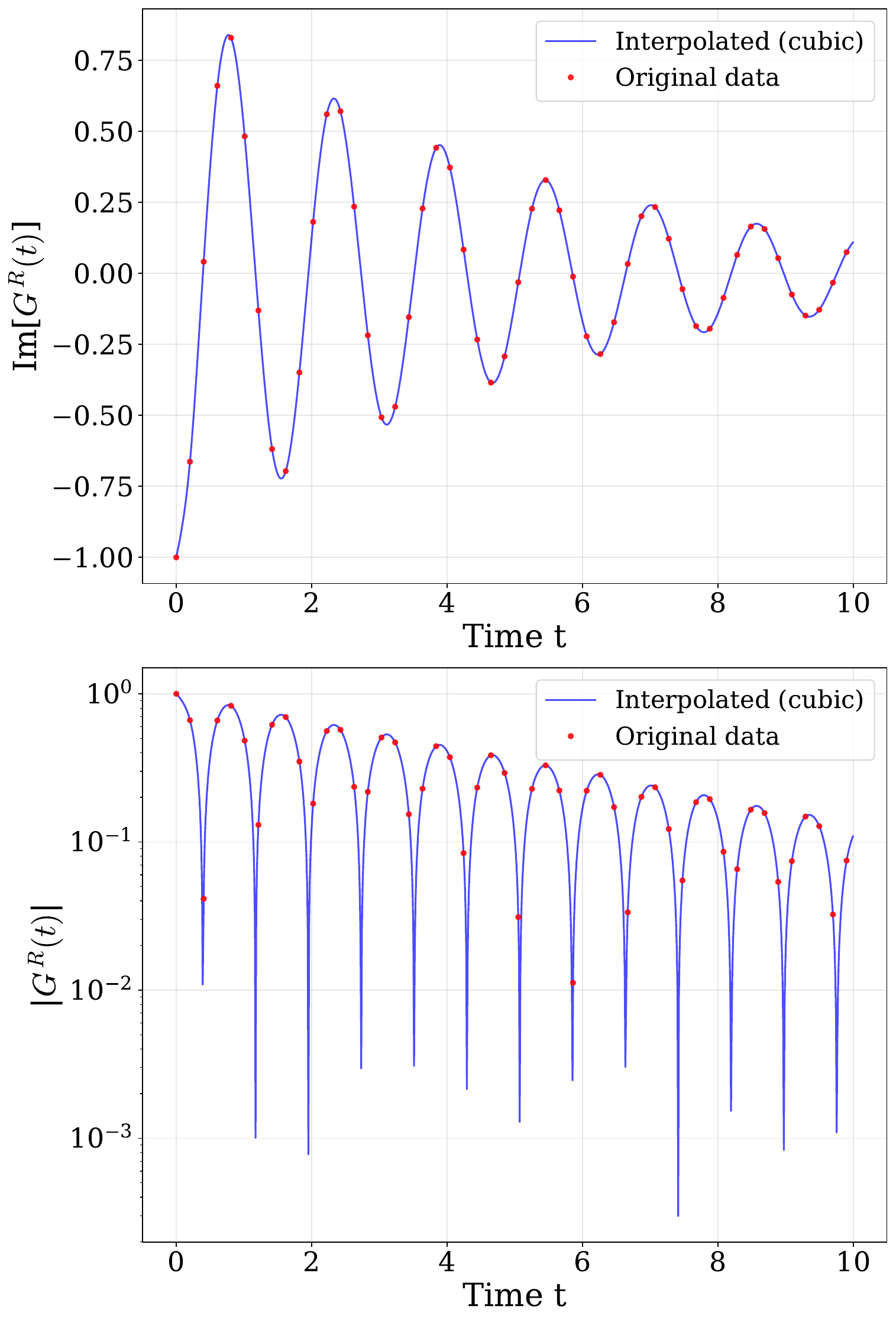}
        \caption{$U=8$}
        \label{fig:3d}
    \end{subfigure}
    
    \caption{Interpolation of the converged impurity Green's function $G^R_{imp}(t)$ across the four interaction strengths ($U = 2, 4, 6, 8$) using cubic splines. \textbf{Top Row:} The imaginary part $\text{Im}[G^R(t)]$ plotted on a linear scale. The red dots represent the discrete time points from the solver, while the solid blue line shows the cubic spline interpolation. \textbf{Bottom Row:} The magnitude $|G^R(t)|$ plotted on a logarithmic scale to highlight the fitting quality at smaller amplitudes and later times. The evolution from slow decay at $U=2$ to rapid, sustained oscillations at $U=8$ reflects the emergence of high-energy Hubbard bands.}
    \label{fig:interpolation}
\end{figure*}

As the interaction strength increases to $U = 4$ [Fig. \ref{fig:2b}], we notice that $t_0$ begins to decrease while the on-site energies $\epsilon_1$ and $\epsilon_2$ increase. This signals a redistribution of spectral weight away from the Fermi level, consistent with the development of correlation-induced features in the spectral function. The convergence remains stable, though larger transient fluctuations are observed during the initial iterations before the parameters settle to their final values.

Around the putative phase transition near $U = 6$ [Fig. \ref{fig:2c}], the value of $t_0$ decreases further, while the magnitude of the on-site energies continues to grow. This behavior reflects the system's approach to the metal-insulator transition, where spectral weight is increasingly transferred from the Fermi level to the emerging Hubbard bands. Notably, the internal bath hoppings $t_1$ and $t_2$ also exhibit significant renormalization, indicating that the effective bath structure must reorganize substantially to capture the strongly correlated physics in this regime. 

Finally, in the Mott insulating regime ($U = 8$) [Fig. \ref{fig:2d}], the parameters depart sharply from their weak-coupling values. The impurity-bath coupling $t_0$ reaches its minimum, effectively decoupling the impurity from the low-energy excitations, while the magnitudes on-site energies $\epsilon_1$ and $\epsilon_2$ are maximized. This parametrization effectively constructs a ``gapped" bath spectrum, faithfully reproducing the charge gap characteristic of the Mott insulator. Despite this significant reorganization of the bath parameters, the iteration scheme maintains robust stability and convergence.

The fact that convergence is consistently achieved within approximately 20 iterations across all interaction strengths is a significant computational advantage. This efficiency is particularly critical for future quantum computing implementations, where each iteration incurs substantial overhead from state preparation and measurement. Furthermore, the linear mixing strategy ($\alpha=0.8$) employed in our algorithm proves effective, striking a robust balance between convergence speed and numerical stability.

\subsection{Retarded Green's Function in Time}

Figure~\ref{fig:interpolation} assesses the quality of the converged impurity Green's function, $G^R_{imp}(t)$, across the same four representative interaction strengths as before. To provide a comprehensive view of the data quality, each panel presents two complementary plots: the upper plot displays the function on a linear scale, while the lower plot uses a logarithmic scale to scrutinize the agreement at smaller amplitudes and longer times. A critical component of our post-processing pipeline is the application of cubic spline interpolation to the raw discrete data. As shown in the figure, the spline generates a smooth, continuous representation from the relatively coarse time grid ($N_t=100$). This step is essential for obtaining accurate spectral functions. Given the limited temporal resolution, a direct discrete Fourier transform would introduce spurious high-frequency oscillations and spectral leakage. The cubic spline enforces continuity in the first and second derivatives at the grid points, effectively filtering these numerical artifacts and ensuring a clean transformation to the frequency domain.

It is important to acknowledge that in the regime of very weak interactions ($U \lesssim 1$), our approximation using only five bath sites encounters inherent limitations. In this limit, the exact spectral function is characterized by a smooth, continuous density of states near the Fermi energy, characteristic of a nearly non-interacting metal. However, the discrete nature of our finite bath leads to a sparse spectrum, resulting in insufficient resolution to capture the fine structure of these low-energy excitations. Specifically, finite-size effects manifest as artificial gaps or discretization steps in the spectral function at energy scales below the typical level spacing $\delta E \sim W/N_{bath}$, where $W$ is the effective bandwidth. 

Consequently, for extremely weak interactions, this method may yield quantitatively inaccurate results due to the inability to reproduce the continuum limit. This limitation becomes less severe at $U = 2$, where correlation effects begin to redistribute spectral weight away from the Fermi level, rendering the discrete representation more adequate. For stronger interactions ($U \geq 4$), this issue is progressively mitigated as the relevant physics becomes dominated by well-separated energy scales, specifically the high-energy Hubbard bands, which are amenable to representation by a small number of discrete poles.

For the intermediate interaction strength ($U = 4$), the Green's function develops a richer temporal structure. We observe enhanced oscillations and a faster decay envelope compared to the weak-coupling case, reflecting the growing electron correlations and the initial redistribution of spectral weight from the quasiparticle peak to high-energy satellites. Approaching the metal-insulator transition at $U = 6$, the impurity Green's function exhibits pronounced oscillations on significantly shorter time scales. These rapid temporal variations are the time-domain signature of the nascent Hubbard bands forming at energies $\omega \approx \pm U/2$. The fitting procedure demonstrates high accuracy in the short-time regime ($t < 5$), which is critical as this window encodes the high-energy spectral features. The logarithmic scale reveals slightly larger deviations at intermediate times compared to weaker interactions; this is an expected consequence of the system entering a critical regime where the simultaneous presence of a narrowing quasiparticle peak and broadening Hubbard bands challenges the resolution of the sparse time grid.

In the Mott insulating regime ($U = 8$), the dynamics are dominated by rapid oscillations starting from $t=0$, corresponding to the fully formed high-energy Hubbard bands. Despite the inherent constraints of using only five bath sites, the fitted Green's function faithfully captures the essential physics of the Mott state. Crucially, the discrete bath representation becomes increasingly effective in this strong-coupling limit, as the spectral weight concentrates into well-separated Hubbard bands that are naturally amenable to representation by a clustered set of discrete bath poles.

Overall, Fig.~\ref{fig:interpolation} confirms that the 5-site bath parametrization successfully reproduces the self-consistent impurity Green's function across the entire crossover from the correlated metal to the Mott insulator. 
This validates our strategy of using a minimal finite bath: the method proves robust in the intermediate-to-strong correlation regime ($U \geq 2$), where the relevant high-energy physics can be adequately resolved even with a small number of discrete degrees of freedom.

\subsection{Spectral Functions and Self-Energy}

\begin{figure*}[t] 
    \centering
    \begin{subfigure}[b]{0.24\textwidth} 
        \centering
        \includegraphics[width=\textwidth]{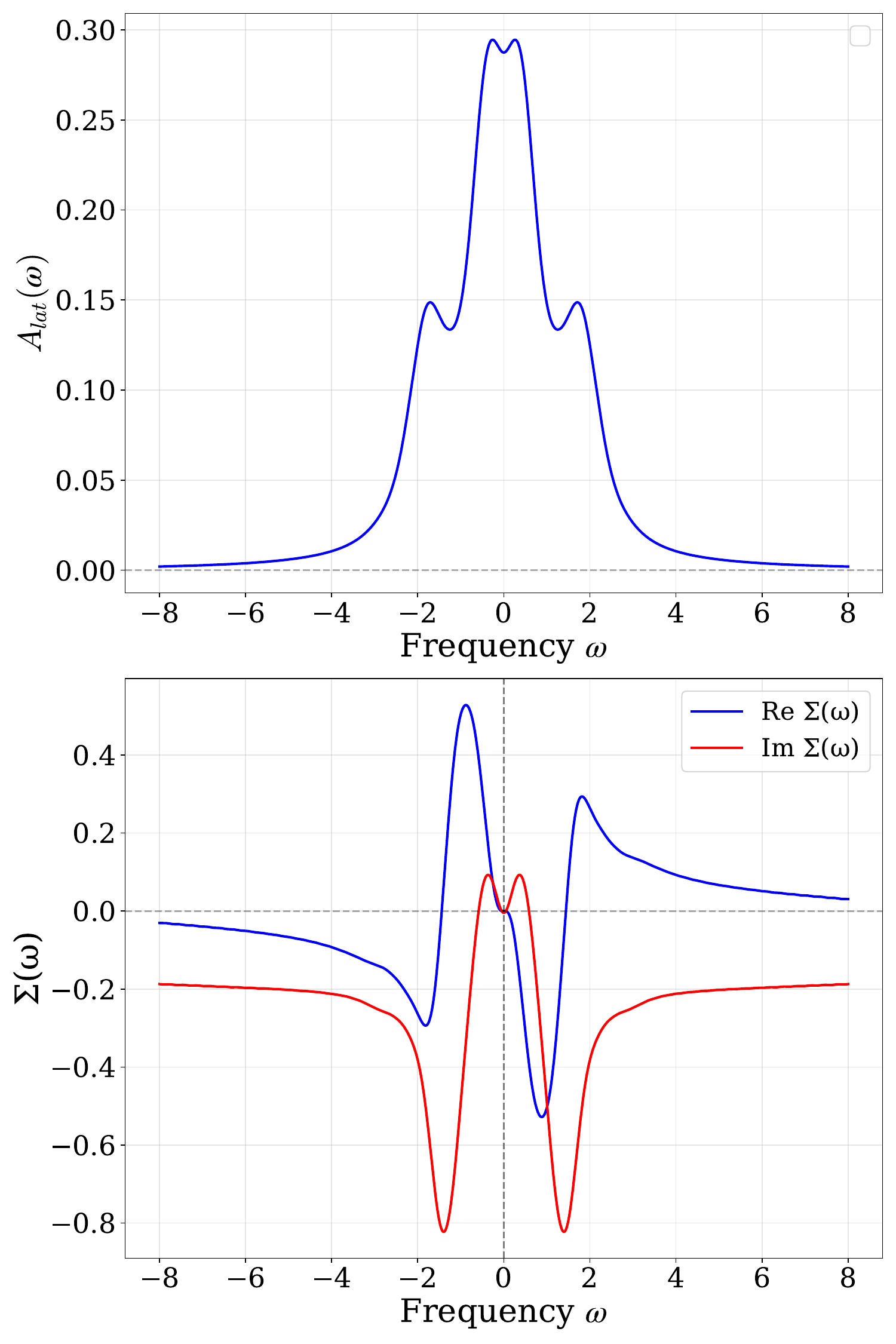}
        \caption{$U=2$}
        \label{fig:4a}
    \end{subfigure}
    \hfill 
    \begin{subfigure}[b]{0.24\textwidth}
        \centering
        \includegraphics[width=\linewidth]{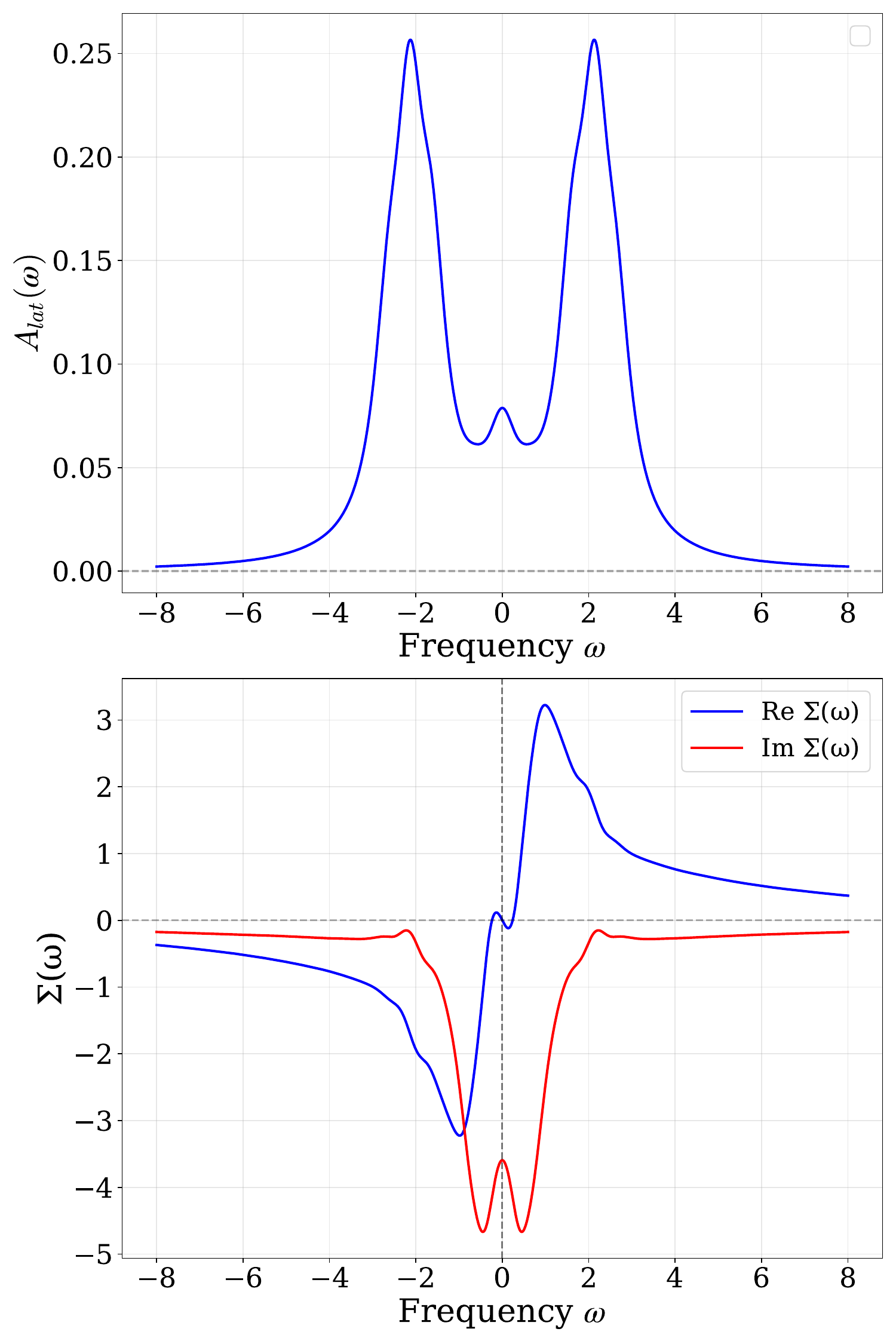}
        \caption{$U=4$}
        \label{fig:4b}
    \end{subfigure}
    \begin{subfigure}[b]{0.24\textwidth}
        \centering
        \includegraphics[width=\linewidth]{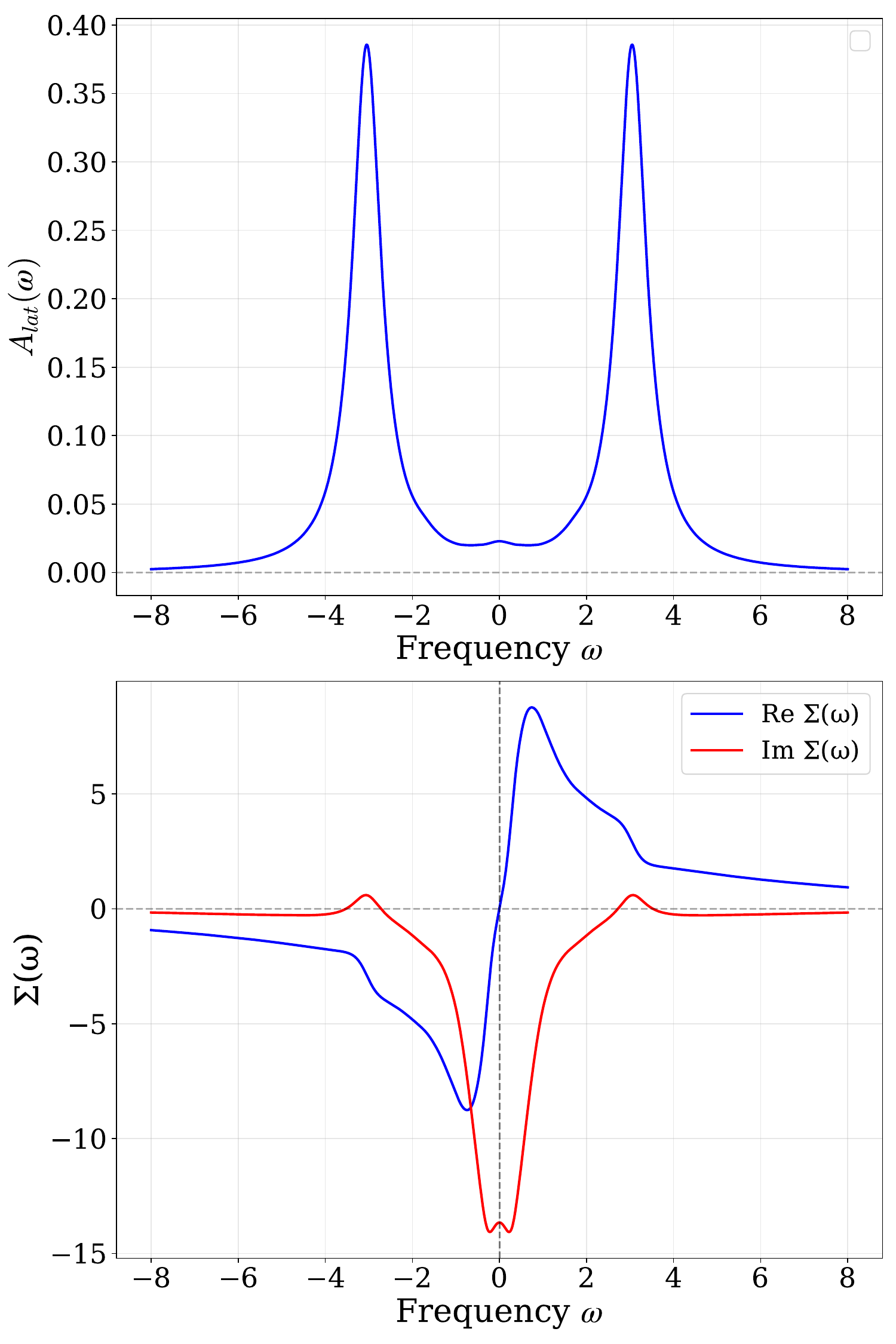}
        \caption{$U=6$}
        \label{fig:4c}
    \end{subfigure}
    \hfill
    \begin{subfigure}[b]{0.24\textwidth}
        \centering
        \includegraphics[width=\linewidth]{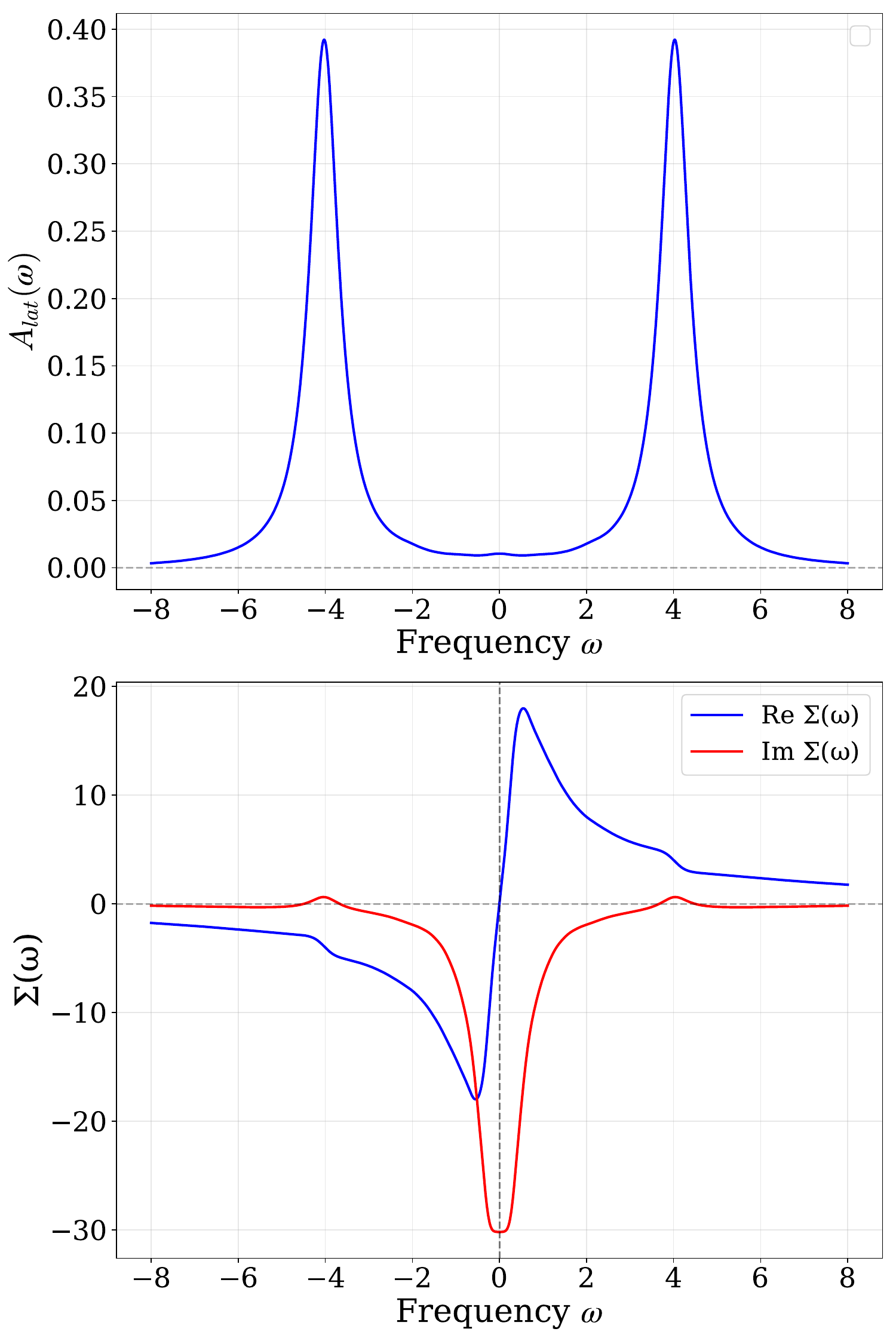}
        \caption{$U=8$}
        \label{fig:4d}
    \end{subfigure}
    
    \caption{Evolution of the single-particle spectral properties across the metal-insulator crossover for interaction strengths $U = 2, 4, 6, 8$.  \textbf{Top Row:} The lattice spectral function $A_{lat}(\omega)$. The system transitions from a correlated metal ($U=2$) with a dominant central quasiparticle peak, through a coexistence regime ($U=4, 6$) characterized by a ``three-peak structure'', to a fully gapped Mott insulator ($U=8$) where spectral weight is  transferred to the high-energy Hubbard bands.
    \textbf{Bottom Row:} The local self-energy $\Sigma(\omega)$, showing the real (blue) and imaginary (red) parts. In the metallic regimes ($U=2, 4$), $\text{Im}[\Sigma(\omega)]$ exhibits a minimum magnitude at $\omega=0$ 
    indicating the presence of long-lived quasiparticles. In the Mott insulating limit ($U=8$), the real part develops a pole-like divergence and the imaginary part diverges at $\omega=0$.}
    \label{fig:spectrum}
\end{figure*}

The central quantity of interest in our analysis is the single-particle spectral function, $A(\omega) = -\frac{1}{\pi} \text{Im} G_{loc}^R(\omega)$. Since our iteration scheme operates entirely in the real-time domain, we must perform a numerical Fourier transform to extract frequency-dependent features. This process presents two primary numerical challenges arising from the limited resources of the simulation: the finite duration of the time evolution ($t_{max}$) and the sparsity of the time grid ($N_t$).

The first challenge concerns the low-frequency resolution, which is fundamentally bounded by the inverse of the simulation duration ($\delta \omega \sim 1/t_{max}$). While the damping factor $\eta$ effectively suppresses contributions from times $t > t_{max}$, it broadens spectral features, further limiting the resolution at the Fermi energy. Also, it is highly unlikely to achieve a reliable extrapolation over long times. Given that our physical model is already limited by the discrete nature of the 5-site bath, attempting to resolve energy scales finer than the bath level spacing would be physically artificial. We therefore accept the intrinsic resolution limit imposed by the combination of the finite bath, finite $t_{max}$, and damping $\eta$.

The second challenge arises from the sparsity of the time grid, which can introduce aliasing artifacts during the Fourier transform. To mitigate this, we employ the cubic spline interpolation method demonstrated in Fig.~\ref{fig:interpolation}. This procedure maps the raw simulation data onto a dense time grid, ensuring a smooth, continuous function that eliminates high-frequency noise while preserving the physical signal. Note that the function is smooth for $t > 0$, except for the discontinuity at $t=0$ required by the fermionic anti-commutation relations.

Figure ~\ref{fig:spectrum} presents the main results of our study: the lattice spectral function $A(\omega)$ and the self-energy $\Sigma(\omega)$ for the four interaction strengths. The spectral function provides direct information about the single-particle density of states, while the self-energy characterizes the renormalization effects due to electron-electron interactions.

\textbf{Weak Coupling ($U=2$):} The spectral function in Fig.~\hyperref[fig:spectrum]{4(a)} exhibits a broad metallic distribution centered at $\omega=0$. While the envelope is consistent with the expected semi-circular density of states of the Bethe lattice ($W = 4t^* = 4$), the profile displays visible deviations and side lobes. These are primarily artifacts of the finite bath approximation ($N_{bath}=5$), which discretizes the continuum into a finite number of poles. The real part of the self-energy, $\text{Re}[\Sigma(\omega)]$, shows the characteristic behavior of a correlated metal, where it crosses zero near the Fermi level with a negative slope, leading to mass renormalization. The imaginary part, $\text{Im}[\Sigma(\omega)]$, exhibits a parabolic minimum near $\omega = 0$. The relatively small magnitude of $\text{Im}[\Sigma(\omega)]$ at the Fermi level indicates that low-energy excitations have long lifetimes and behave as well-defined quasiparticles.

\textbf{Intermediate Coupling ($U=4$):} As the interaction increases to $U=4$ (Fig.~\hyperref[fig:spectrum]{4(b)}), the system enters the distinct coexistence of the central peak and the Hubbard bands. The spectral function separates into a clear three-peak structure, a narrowed central quasiparticle peak at $\omega=0$ and two prominent high-energy lobes at $\omega \approx \pm 2.5$. This is the onset of the lower and upper Hubbard bands, which represent states with reduced or enhanced double occupancy, respectively. The total spectral weight is conserved (as required by sum rules), but its redistribution signals the growing importance of correlation effects. The self-energy at $U = 4$ shows enhanced structure compared to $U = 2$. $\text{Re}[\Sigma(\omega)]$ develops a more pronounced frequency dependence with larger absolute values, indicating stronger renormalization effects. 
The suppression of the scattering magnitude at $\omega=0$ confirms that a coherent metallic state is preserved, even as the scattering rate increases rapidly at higher frequencies corresponding to the Hubbard bands.

\textbf{Near the Transition ($U=6$):} At $U=6$ (Fig.~\hyperref[fig:spectrum]{4(c)}), the system is on the brink of the metal-insulator transition. The spectral function now displays three distinct features: a significantly suppressed central peak near $ \omega= 0$, and two well-separated peaks at $\omega \pm 3$, which are the lower and upper Hubbard bands. The central peak, while still present, has lost most of its spectral weight. This is consistent with the system being near the critical interaction strength $U_c$ for the Mott transition on the Bethe lattice. The positions of the Hubbard bands at $\omega \approx \pm U/2 \approx \pm 3$ align with the expected energy cost for removing or adding an electron to a singly occupied site. The finite width of these bands corresponds to the intrinsic broadening arising from strong correlations and hybridization. Importantly, while the spectral weight at $\omega = 0$ is greatly reduced, it does not vanish completely, suggesting that the system is on the metallic side of the transition but very close to the insulating phase. The self-energy exhibits a very sharp, nearly vertical slope in $\text{Re}[\Sigma(\omega)]$ across the Fermi level. The imaginary part $\text{Im}[\Sigma(\omega)]$ forms a sharp, narrow notch at $\omega=0$. The scattering rate shoots up rapidly just away from the Fermi energy, implying that the quasiparticles are effectively defined only strictly at the Fermi level and are destroyed by strong scattering at any finite excitation energy.

\textbf{Mott Insulator ($U=8$):} For $U=8$ (Fig.~\hyperref[fig:spectrum]{4(d)}), the central peak has vanished completely, leaving a fully gapped spectrum dominated by two symmetric Hubbard bands centered at $\omega\approx\pm4$. The defining signature of the Mott state is clearly visible in the self-energy. The real part $\text{Re}[\Sigma(\omega)]$ exhibits a pole-like divergence at $\omega = 0$. Correspondingly, the imaginary part $\text{Im}[\Sigma(\omega)]$ shows a massive negative spike exactly at $\omega=0$. Physically, this represents an infinite scattering rate at the Fermi level, implying the concept of a propagating quasiparticle breaks down entirely for the insulating state.

\section{Conclusion}\label{Section:Conclusion}
We have demonstrated a method for solving the Hubbard model using dynamical mean-field theory via an iteration in the time-domain. The solver is based on a mapping to a one-dimensional chain with a finite number of bath sites, where the output is the interacting impurity retarded Green's function in real time. Unlike most conventional methods for solving the self-consistent equations of dynamical mean-field theory, which are typically based on imaginary time or frequency, we have demonstrated that reasonably accurate results can be obtained even with a rather small number of bath sites. 

In particular, we obtain the spectral function, which shows clear features of the metal-to-insulator transition as the interaction strength increases—a key feature captured by dynamical mean-field theory. Although we deliberately chose a limited resolution for the calculation of the impurity Green's function, the iteration still yields stable convergence, provided that damping is included in the retarded Green's function. 

Unlike most calculations related to solving the dynamical mean field theory via quantum computing approach to date, this approach allows us to obtain more detailed features of the spectral function, rather than merely capturing the quasi-particle weight as in the so-called two-site dynamical mean-field theory approximation \cite{Ayral2025,Pothoff1999}. An obvious next step is to replace the exact diagonalization solver used here with a quantum computing-based solver. Since it is unlikely, at least in the short term, that quantum hardware will allow us to directly extract the properties of strongly correlated materials, approximations must be employed to make progress; dynamical mean-field theory combined with density functional theory represents a strong candidate to fill this role. The present study suggests that such calculations may be feasible beyond perturbative solvers or the two-site approximation. Of course, developing a robust impurity solver that is feasible on quantum hardware for a moderate number of bath sites remains a nontrivial problem in itself.

The framework established here paves the way for several promising extensions. First, it is natural to explore the possibility of related approaches for non-equilibrium DMFT, allowing for the direct simulation of pump-probe experiments and transient dynamics \cite{Freericks_Turkowski_Zlatic_2006,Freericks_2008,Aoki_etal_2014,Kreula_Clark_Jaksch_2016}. Second, the computational efficiency gained by using small bath clusters suggests that this method could be possibly adapted for Extended DMFT to treat non-local interactions \cite{Backes_etal_2023}. Furthermore, combining this real-time approach with Real-Space DMFT would open new avenues for studying inhomogeneous systems \cite{Freericks_2004,Eckstein_Werner_2013}. An encouraging sign is that the VQE method appears remarkably resilient to random disorder \cite{Alvertis_etal_2025}, specifically the critical interplay between disorder and interaction could be studied by combining with the coherent potential approximation and related methods \cite{Dohner_etal_2022,Abuelmaged_etal_2025,Rangi_Moreno_Tam_2024,Rangi2025Disorder, Yan_Werner_2023,Soven_1967,Velicky_Kirkpatrick_Ehrenreich_1968}. Finally, the formalism could also be generalized to non-Hermitian Hamiltonians, offering a viable pathway to investigate the spectral properties of open quantum systems or effective models exhibiting gain and loss \cite{Rangi_Moreno_Tam_2025,Xie_Xue_Zhang_2024}. 

\begin{acknowledgments}
This work used high-performance computational resources provided by the Louisiana Optical Network Initiative and HPC@LSU computing.
\end{acknowledgments}

\section*{Data Availability}

The data that support the findings of this article are openly available.

\bibliography{refs}
\newpage
\appendix
\onecolumngrid
\end{document}